\documentclass{elsart}
\usepackage{graphicx,amssymb}
\def\be{\begin{equation}}
\def\ee{\end{equation}}
\def\ba{\begin{eqnarray}}
\def\ea{\end{eqnarray}}
\journal{Physica A}

\begin{document}
\begin{frontmatter}
\title{Nearest neighbour spacing distribution of basis in some intron-less and 
intron-containing DNA sequences}

\author{M.~F.~Higareda}
\address{Instituto de Investigaciones Biom\'edicas, 
Universidad Nacional Aut\'onoma de M\'exico, 
Apdo. Postal 70228, Ciudad Universitaria, 
C.P. 04510 M\'exico D.F. M\'exico}

\author{H.~Hern\'andez-Salda\~na}
\address{Universidad Aut\'onoma Metropolitana-Azcapotzalco, 
M\'exico D.F., M\'exico and \\
Instituto de F\'{\i}sica, Universidad Nacional
Aut\'onoma de M\'exico, M\'exico D.F, M\'exico}

\author{R.~A.~M\'endez-S\'anchez}
\address{Centro de Ciencias F\'{\i}sicas, 
Universidad Nacional Aut\'onoma de M\'exico, 
A.P 48-3, 62210, Cuernavaca, Morelos, M\'exico}


\begin{abstract}
We show that the nearest neighbour distribution of distances 
between  basis pairs  of  
some intron-less and intron-containing coding regions 
are the same when a procedure, called {\em unfolding}, is 
applied. Such a procedure consists in separating the secular variations
from the oscillatory terms. The form of the distribution obtained 
is quite similar to 
that of a random, i.e. Poissonian, sequence. This is done 
for the HUMBMYH7CD, DROMYONMA, HUMBMYH7 and DROMHC sequences. 
The first two correspond to highly coding regions 
while the last two correspond to non-coding regions.
We also show that the distributions before 
the unfolding procedure depend on the secular part 
but, after the unfolding procedure we obtain an striking result: 
all distributions are similar to each other. The result becomes independent
of the content of introns or the species we have chosen.
This is in contradiction with the results obtained with 
the detrended fluctuation analysis in which the correlations 
yield different results for intron-less and intron-containing regions.
\end{abstract}

\begin{keyword}
Spectral statistics, genomic sequences, statistical analysis
\PACS 87.14.Gg, 87.10.+e, 05.40.-a, 87.15.Ya
\end{keyword}
\date{.}
\end{frontmatter}

\section{Introduction}
\label{Sec:Introduction}
In recent times DNA of new species have been sequenced, opening
new questions in biological physics. 
Large data of genomic sequences
are now available and ready to be analyzed. 
In particular, the study of their statistical properties 
is of broad interest. Up to now, several works addressed 
the correlations in coding and non-coding regions~\cite{LiKaneko,Pengetal-1}. 
Many statistical measures have been introduced since those works 
appeared: the detrended 
fluctuation analysis~\cite{Pengetal-2}, diffusion 
entropy analysis~\cite{ScafettaLatoraGrigolini}, factorial moment 
analysis~\cite{MohantyNarayana-Rao}, wavelet 
analysis~\cite{ArneodoBacryGravesMuzy}, among others. 

Despite several efforts (see Ref.~\cite{MohantyNarayana-Rao} and 
references therein), the question about the statistical properties 
between intron-less and intron-containing coding regions is still open. 
Here we use two complementary techniques to study the 
``short-range'' statistical properties of DNA sequences. Our work considers,
at this stage, only monoplets, but studies on duplets and triples are  in 
progress ~\cite{HigaredaHernandez-SaldanaMendez-Sanchez}.
First we use a generalized detrended fluctuation analysis, commonly called 
{\em unfolding}~\cite{Metha,Brodyetal,GuhrMuellerGroelingWeidenmueller}, 
to separate the secular trend and the non-secular properties 
of the sequences. Then, the nearest neighbour distribution is obtained 
and analyzed.

In the next section we define the cumulative basis density and 
the mathematical procedure called {\em unfolding}. The cumulative basis 
density is obtained for: two intron-less sequences (0\% of introns), 
HUMBMYH7CD and DROMYONMA, corresponding to the Human 
$\beta$-cardiac myosin heavy chain (MHC) 
and {\em Drosophila melanogaster} MHC, respectively; 
and two intron-containing coding sequences, HUMBMYH7 and DROMHC that 
correspond to the Human $\beta$-cardiac MHC and to the {\em 
Drosophila melanogaster} MHC, respectively. The sequences 
were taken mainly from GenBank\cite{GenBank} and were analyzed with the 
detrended fluctuation analysis \cite{Pengetal-1}. 
The number of introns of each sequence is given in 
Table~\ref{Table:Sequences}.
In section~\ref{Sec:NearestNeighbor} the nearest neigbour distribution 
is obtained yielding a universal feature.
Examples of the nonsense results obtained without a proper unfolding procedure 
are also given there. A brief conclusion follows.

\begin{table}
\begin{tabular}{lcccccccccc}
\hline
code & size & introns & \multicolumn{4}{c}{basis-content (\%)} & 
\multicolumn{4}{c}{$\langle s \rangle $ (basis)}\\
&(basis)& (\%) & A & T & C & G & A & T & C & G  \\
\hline
HUMBMYH7CD & 5999  &   0 & 26.7 & 16.0 & 26.0 & 31.3 & 3.7 & 6.3 & 3.8 & 3.2 \\
DROMYONMA  & 6338  &   0 & 24.1 & 14.5 & 18.7 & 21.7 & 3.3 & 5.4 & 4.2 & 3.6    \\
HUMBMYH7   & 28438 &  73 & 23.6 & 23.0 & 27.4 & 26.0 & 4.2 & 4.4 & 3.6 & 3.9   \\
DROMHC     & 22663 &  72 & 30.6 & 18.4 & 23.6 & 27.5 & 3.3 & 5.5 & 4.3 & 3.7 \\
\hline
\end{tabular}
\caption{Some statistical properties of the sequences analyzed here. 
A, T, C, and G correspond to adenine, thymine,
cytosine and guanine, respectively.
$\langle s \rangle $ stands for the mean level spacing of each basis and 
is calculated as an arithmetic average.}
\label{Table:Sequences}
\end{table}

\section{Unfolding procedure}
\label{Sec:UnfoldingProcedure}

Following Ref.~\cite{MohantyNarayana-Rao} we define the 
density of basis as 
\be
\rho^\beta (x)=\sum_{i=1}^{N_\beta} \delta(x-x_i^\beta)
\ee
where $\{ x_i^\beta \}_{i=1}^{N_\beta}$ is the ordered set of
$N_\beta$ positions of the basis. Here $\beta$ stands for
adenine (A), thymine (T), cytosine (C), or
guanine (G) in DNA. In the last equation $\delta(x)$ is the 
Dirac $\delta$-function.
The cumulative basis density or staircase function is defined 
as the derivative of $\rho^\beta (x)$:
\be
\mathrm{N}^\beta (x)=\sum_{i=1}^{N_\beta} \Theta(x-x_i^\beta),
\ee
where $\Theta(x)$ is the Heaviside function. 
This function is plotted in Fig.~1 for the sequences 
HUMBMYH7CD, DROMYONMA, HUMBMYH7 and DROMHC.
Notice that this quantity corresponds to the inverse function 
of the cumulative position plots of Ref.~\cite{SanchezJose}. 

\begin{figure}
\begin{center}
\includegraphics[width=\columnwidth]{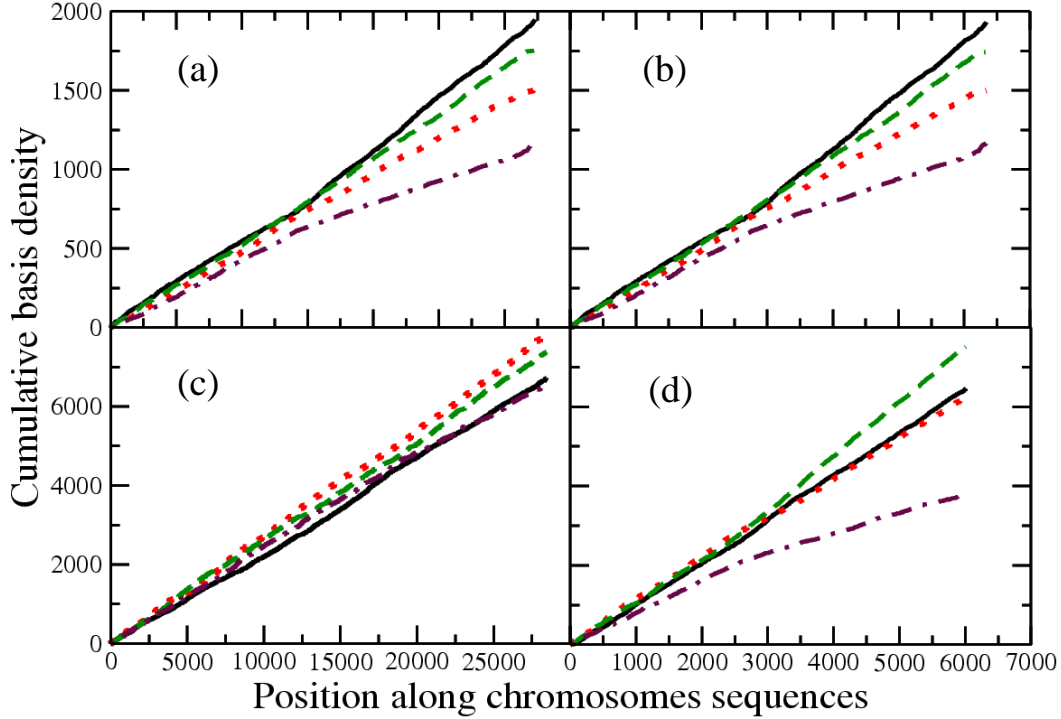}
\caption{(color online) Cumulative basis density of the sequences 
(a) DROMYONMA, (b) DROMHC, (c) HUMBMYH7CD and (d) HUMBMYH7 . The first column 
corresponds to non coding region and the second one to coding one. The 
first row corresponds to {\em D. melanogaster} and the second one to human. 
The solid (black) line corresponds 
to adenine (A), the dash-dotted (maroon) line to thymine (T), the 
dotted (red) line to 
cytosine (C) and the dashed line (green) to guanine (G).}
\end{center}
\label{Fig:N(E)}
\end{figure}

Following Refs.~\cite{Metha,Brodyetal,GuhrMuellerGroelingWeidenmueller}, 
the cumulative basis density can be expressed as 
\be
\mathrm{N}^\beta (x)= \langle \mathrm{N}^\beta (x)\rangle +
 \mathrm{N}_{osc}^\beta (x),
\ee
where $\langle \mathrm{N}^\beta (x)\rangle$ is a mean or secular part 
and $\mathrm{N}_{osc}^\beta (x)$ is the oscillating one. 
The brackets in the secular part refer to average in 
position. Notice that the secular part is a function of $x$ and is not 
necessarily smooth. In quantum physical systems, for instance, 
$\langle \mathrm{N}^\beta (x)\rangle$ contains the periodic orbit information,
{\i.e.} the classical one~\cite{GuhrMuellerGroelingWeidenmueller}.
As seen in Fig.~1, cumulative densities of different 
basis and/or sequences have different secular parts.
Some look almost linear while other appear linear by parts.
It is not clear for us the biological origin of the abrupt change in the slope 
of $\langle \mathrm{N}^\beta (x)\rangle$. However, for bacterial genomes, 
similar 
changes in density corresponds to the replication origin~\cite{SanchezJose}. 
Analytical expressions for the secular part are unknown  up to now,
hence we will use some of the numerical methods commonly used in 
spectral statistics. For instance, we will use a polynomial fitting. 
The degree of the polynomial is chosen, as usual, as the minimum in which 
the fluctuations does not change anymore. Notice that this unfolding 
procedure is more general that the detrended fluctuation analysis, 
since the latter uses only a linear fitting. A systematical study and a 
comparison with a different unfolding procedure (window analysis)
will be given elsewhere~\cite{HigaredaHernandez-SaldanaMendez-Sanchez}.
In the sequences of Table~\ref{Table:Sequences} we tested fittings 
with polynomials from 8th until 18th degree. We also used windows 
from 100 to 1000 distances. Within this range of parameters 
(polynomial degree and window) the results we present are stationary, 
i.e. the results do not change.

The transformation that allows the comparison of the fluctuations for systems 
with different secular parts 
is the unfolding. It works as follows. Lets define a new 
unfolded sequence $\{ y_i^\beta \}_{i=1}^{N_\beta}$ where 
\be
y_i^\beta=\langle \mathrm{N}^\beta (x_i^\beta) \rangle.
\ee
With this transformation the new sequence has a uniform density and mean 
level spacing equal to unity, {\it i.e.} 
$\langle y_{i+1}^\beta-y_i^\beta \rangle = 1$.

\section{Nearest neigbour spacing distribution}
\label{Sec:NearestNeighbor}

The nearest neigbour spacing distribution, $p(s)$, is the observable 
most commonly used to study the ``short-range'' fluctuations in the 
density of basis. Here $s_i^\beta=y_{i+1}^\beta-y_i^\beta$.
The quotes mean that the short range is defined in 
units of the mean level spacing $\langle s^\beta \rangle$.  
Then, for a very diluted basis in a sequence,  
$p(s)$ refers to very long distances in the complete sequence and vice versa. 
The distribution $p(s)$ then yields the probability density to have two 
neigbouring levels $y_i^\beta$ and $y_{i+1}^\beta$ at a spacing $s$.
With the unfolding procedure the function $p(s)$ and its first moment 
are normalized to unity.  The mean level spacing $\langle s \rangle$ on  
basis, without unfolding,   
is given in Table~\ref{Table:Sequences}. Notice that this is not a 
well defined quantity for some sequences like  thymine in DROMYONMA 
(dash-dotted in Fig.~1), since it has two 
different values and hence, the unfolding procedure is required. 

\begin{figure}
\includegraphics[width=\columnwidth]{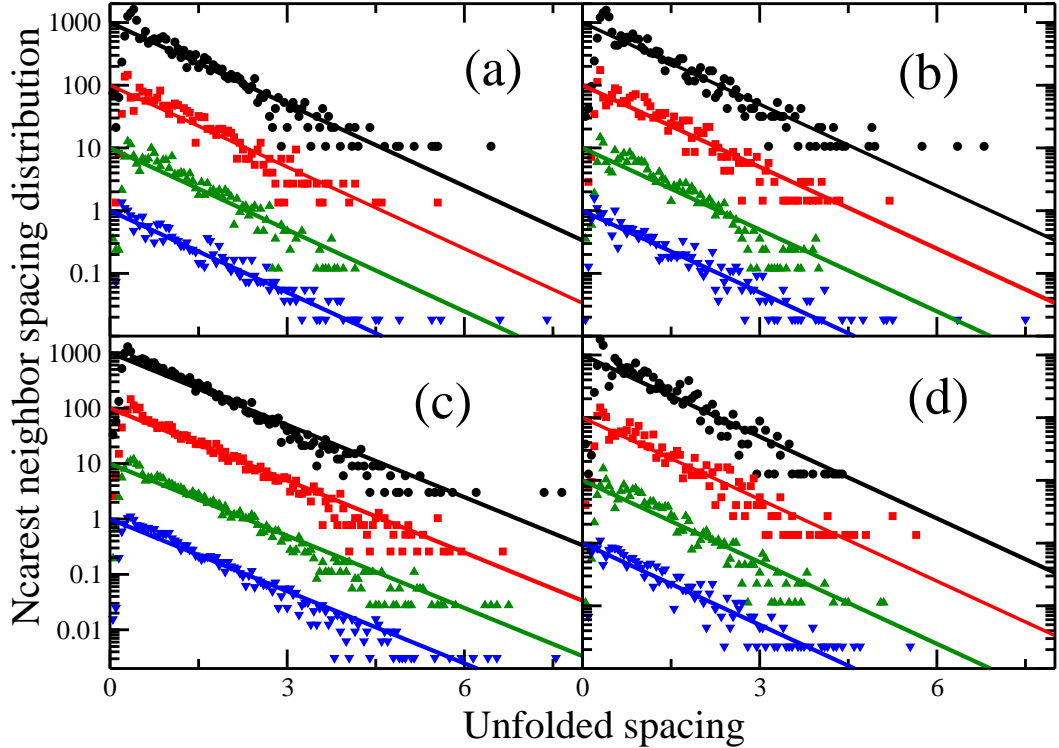}
\caption{(color online) Nearest neigbour spacing distribution, $p(s)$, 
for the same sequences of Fig.~1 and following the 
same order. The circles (black) correspond to the $p(s)\times 1000$ of 
adenine, the squares (red) to the $p(s)\times 100$ of cytosine,
the triangles up to the $p(s)\times 10$ of guanine and the triangles 
down (blue) to the $p(s)$ of thymine. The solid lines correspond to 
the Poissonian result $p(s)=\exp(-x)$ times the corresponding scaling factor.}
\label{Fig:p(s)}
\end{figure}

In Fig.~\ref{Fig:p(s)} we show the obtained nearest neigbour spacing 
distribution $p(s)$ for the different basis for the different sequences.
As seen in this figure, they are very similar to the 
uncorrelated Poissonian prediction 
\be
p(s)=exp(-s).
\ee
The main differences between the results of the genomic sequences and 
the Poissonian prediction appear at both, very short and 
very long distances. The explanation at short distances is trivial: 
they are  related to the physical size of the basis which avoid  
distances smaller than the physical size of a basis. This generates 
a ``repulsion hole''. The repulsion hole is not due to the unfolding 
procedure but this procedure makes the hole much more evident.
The result we get in \ref{Fig:p(s)} is opposite to the well-known 
result given in \cite{Pengetal-2}.
This means that, at least in the sequences we used, it is impossible 
to differentiate coding and non-coding regions with statistical analysis. 
Biological information is present only in the secular part but not in 
the fluctuations.

\begin{figure}
\includegraphics[width=\columnwidth]{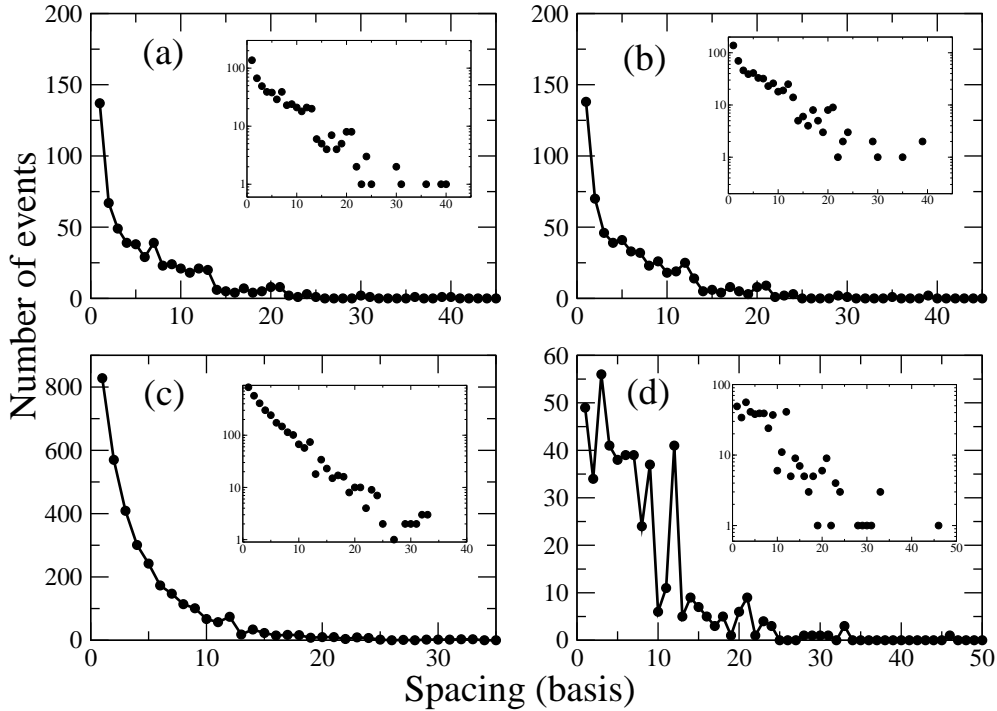}
\caption{Nearest neigbour spacing histograms for thymine in the same 
order as the previous figure. No normalization nor unfolding was performed. 
Notice that the fluctuation at small spacings are large in case (a) and (b) 
and huge in case (d). An exponential fitting in case (c) is
 correct however it is non-sense in case (d). The insets show the same 
plots with the Y-axis in logaritmic scale.}
\label{Fig:WRONGp(s)}
\end{figure}

Now we will show some results without the unfolding procedure. 
In Fig.~\ref{Fig:WRONGp(s)} we show the histograms we obtained for thymine 
in the four sequences. They 
are no normalized nor unfolded. 
As seen, some of them, Fig.~\ref{Fig:WRONGp(s)} (a) and (c), are very 
similar to the Poissonian prediction while other, 
Fig.~\ref{Fig:WRONGp(s)} (b) and (d), are quite different. This behavior give
the impression that statistical distribution for non coding and coding regions 
are different. However this wrong conclusion is due to an artifact 
because in Fig.~\ref{Fig:p(s)} they have clearly the same distribution. 
The explanation of this behavior is very simple:
since the density is not uniform and it depends on the position along the 
sequences, the distances between basis have different statistical weights.
Missing or incorrect rectification (unfolding), gives place to an incorrect 
measurement of the fluctuations and to possible wrong conclusions. 
Then, the unfolding procedure eliminates artifacts due to a wrong average.
To find the correct way to unfold a sequence is a very 
delicate and difficult task~\cite{GuhrMuellerGroelingWeidenmueller}.
However, it is the way to obtain a measure of the fluctuations which 
allow us a comparison between systems having different trends.

\section{Conclusions}

We introduced the cumulative basis density as well as the unfolding 
procedure to analyze distances between basis in genomic sequences.
We have shown that the nearest neigbour spacing distribution 
is very close to the Poissonian distribution. The main differences 
between them have been found in short distances where the 
physical size of each 
basis generates a ``repulsion hole''. The results we have got for the 
nearest neigbour spacing distribution are quite different that those 
available in the literature for other quantities like correlations. 
We obtained the same result, almost Poissonian, for two 
intron-less and two intron-containing coding sequences. 
This suggest that a proper unfolding procedure should be implemented before 
any statistical analysis of genomic sequences. Artifacts due to 
a wrong average could appear. Our results indicate strongly that 
statistical measures in genomic sequences without proper 
unfolding could give artifacts and probably several results in the 
literature on the area are not correct. 

\section*{Acknowledgements}
It is a pleasure for us to deeply thank to MV Jos\'e and the theoretical 
biology group at IIB-UNAM for their inestimable support and great help.
We wish to thank to C. Abreu-Goodberg  for helpful discussions and 
E. V\'arguez-Villanueva for useful comments.
This work was partially supported by CONACyT and DGAPA-UNAM IN104400.
MFH thanks CONACyT by a Ph. D. grant. With this paper we wish to celebrate 
with Professor Alberto Robledo his 60th anniversary and to have the 
opportunity to do physics at mesolatitudes.
Alberto, not only happy birthday but that all non liquid works be fluid.

\end{document}